# Highly Conducting Π–Conjugated Molecular Junctions Covalently Bonded to Gold Electrodes

Wenbo Chen,[†] Jonathan R. Widawsky,[‡] Héctor Vázquez,[‡] Severin T. Schneebeli,[†] Mark S. Hybertsen*,[§] Ronald Breslow*,[†] and Latha Venkataraman*[‡]

Departments of [†]Chemistry and [‡]Applied Physics & Applied Mathematics, Columbia University, New York, New York 10027 and Center for Functional Nanomaterials,[§] Brookhaven National Laboratory, Upton, New York 11973-5000

**Supporting information placeholder**

**ABSTRACT:** We measure electronic conductance through single conjugated molecules bonded to Au metal electrodes with direct Au-C covalent bonds using the scanning tunneling microscope based break-junction technique. We start with molecules terminated with trimethyltin end groups that cleave off *in situ* resulting in formation of a direct covalent sigma bond between the carbon backbone and the gold metal electrodes. The molecular carbon backbone used in this study consist of a conjugated π system that has one terminal methylene group on each end, which bonds to the electrodes, achieving large electronic coupling of the electrodes to the π system. The junctions formed with the prototypical example of 1,4-dimethylenebenzene show a conductance approaching one conductance quantum ($G_0 = 2e^2/h$). Junctions formed with methylene terminated oligophenyls with two to four phenyl units show a hundred-fold increase in conductance compared with junctions formed with amine-linked oligophenyls. The conduction mechanism for these longer oligophenyls is tunneling as they exhibit an exponential dependence of conductance with oligomer length. In addition, density functional theory based calculations for the Au-xylylene-Au junction show near-resonant transmission with a cross-over to tunneling for the longer oligomers.

It is a great challenge to achieve electronically transparent connections between metal electrodes and organic molecules,[1] so as to minimize resistances introduced by the chemical linkers normally used to form such interfaces.[2] Typically, thiols[2b,3] that bind covalently to gold, or amines[2a,2b,4] that form donor-acceptor bonds to under-coordinated gold, are used to electronically couple organic backbones to metal electrodes. For each link group, analysis of a series of single molecule junctions as a function of length has generally revealed a large contact resistance, significantly larger than the ideal limit for a single channel, $1G_0$.[2a,2c,5] A junction with conductance close to $G_0$ has been demonstrated for $H_2$ and benzene molecules with platinum electrodes under high vacuum conditions at low temperatures.[2d,6] However, the ability to create and control transport through highly conducting molecular-metal interfaces still remains a major challenge, especially under ambient conditions.

We have shown previously that direct Au-C covalent sigma bonds can be created *in situ* at the molecule-gold interface, resulting in highly conducting sigma-bonded systems.[7] For example, a conductance of 0.1 $G_0$ through a butane backbone was demonstrated. These direct Au-C bonded molecular junctions were created starting with trimethyltin-terminated alkanes. The trimethyltin end-groups cleaved off *in situ*, yielding direct Au-C bond coupled junctions. Our Density Functional Theory (DFT) based calculations showed that in the limit of a single methylene group a conductance approaching $G_0$ could be achieved, suggesting that the direct Au-C link has near ideal transmission characteristics. While we succeeded in forming junctions with benzene, the conductance was relatively low, consistent with calculations indicating conduction occurred through the sigma system.[7-8]

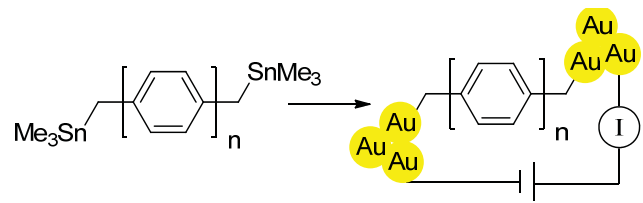

**Figure 1.** *In situ* formation of direct Au-electrode-C bonds starting from SnMe$_3$ precursors. **P1**, n=1; **P2**, n=2; **P3**, n=3; **P4**, n=4

Here, we create single molecule junctions using conjugated backbones terminated with methylene groups that bind covalently to gold metal electrodes, again through the use of SnMe$_3$ groups that cleave off *in situ*. We find that the resulting junctions have a conductance that is a hundred-fold higher than similar junctions formed with conventional linkers.[9] These junctions are highly conducting because the Au-C bonds to the terminal methylene units are well coupled to the conjugated π system. This is in contrast to similar junctions created previously where Au was bound directly to a carbon on the benzene ring.[7] Specifically, we find that the conductance of p-xylylene bonded to gold electrodes approaches 1 $G_0$. Our theoretical calculations show that conductance occurs via near-resonant transmission. For longer polyphenyls with 2-4 phenyl units, we find that the conductance decreases exponentially with increasing length with a characteristic decay constant of 1.9/phenyl group.

We synthesized a series of trimethylstannylmethyl-terminated polyphenyls and measured the conductance of single-molecule junctions formed from these molecules using the scanning tunneling microscope based break-junction (STM-BJ) method (Figure 1).[2a,3b] In this technique, single molecule junctions are created by repeatedly forming and breaking Au point contacts[3b] in a ~10 mM 1,2,4-trichlorobenzene solution of the target trimethyltin-terminated molecules. Conductance

(current/voltage) is measured as a function of the relative tip/sample displacement to yield conductance traces, which are used to generate conductance histograms. We synthesized 1,4-trimethylstannyl terminated xylylene using two different methods. In one, we converted 1,4-bis-bromomethylbenzene (p-xylylene dibromide) to the dilithio compound and reacted it with trimethylstannyl chloride[10]. In the other method, we reacted the p-xylylene dibromide with trimethylstannyl lithium[11]. The latter procedure was used to attach the trimethylstannyl groups for the polyphenyl compounds. The details of these synthetic procedures and characterizations are given in the supporting information document (SI). We note here that these compounds are toxic and should be handled with care.

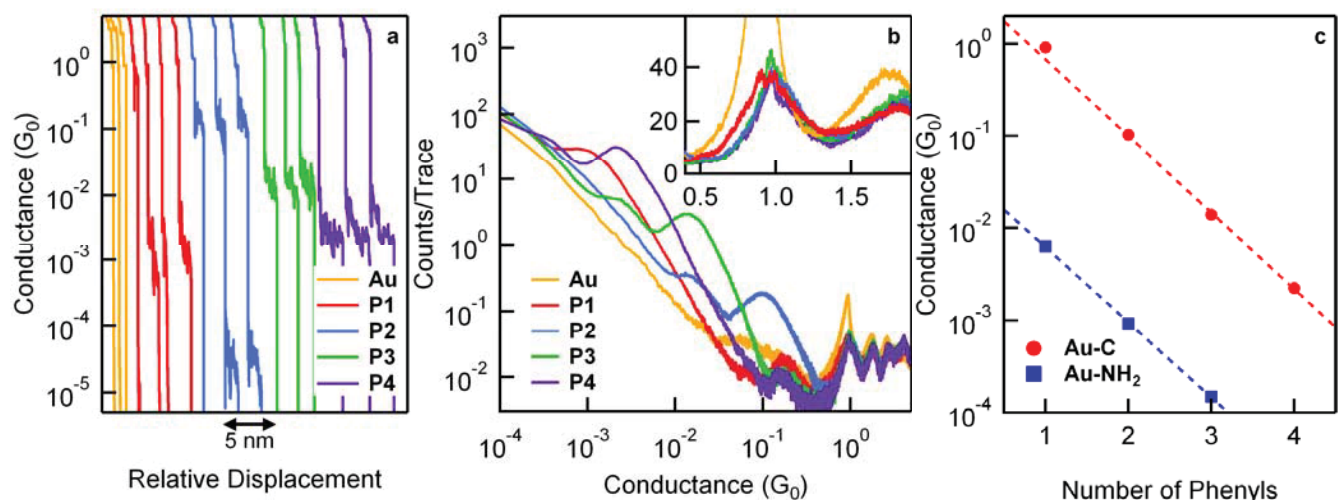

**Figure 2.** (a) Individual conductance traces measured in solutions of the SnMe$_3$ terminated polyphenyl compounds **P1-P4**. Measurements in solvent alone are also shown for comparison (**Au**). The applied bias is 250 mV. (b) Conductance histograms of over 10000 traces generated with linear bin size of 0.0001 G$_0$ shown on a log-log scale. The inset shows the same data on a linear scale. (c) Conductance versus number of phenylene units in the chain for compounds **P1-P4** and for analogous measurements with a diamine series taken from reference 12. Dotted lines represent linear least squares fit to **P2-P4** series. Note the point for **P1** above the line.

Figure 2a compares individual conductance traces from measurements of solutions of stannylated 1,4-dimethylenebenzene, 4,4'-dimethylenebiphenyl, 4,4''-dimethylene-p-terphenyl and 4,4'''-dimethylene-p-tetraphenyl (**P1**, **P2**, **P3** and **P4** respectively). We see clear conductance plateaus at molecule-dependent conductance values, although in the case of **P1** it is not straightforward to distinguish the molecular plateau from that of the single atom contact at a conductance around G$_0$. These plateaus are due to conduction through a molecule bonded in the gap between the two Au point-contacts. These conductance plateaus are seen in the measurements immediately after a solution of the target molecule terminated with SnMe$_3$ groups is added, in contrast with measurements of 1,4-bis(trimethylstannyl)benzene[7], where conductance plateaus were seen in measurements only after 2.5 hrs. This delay seen in measurements of 1,4-bis(trimethylstannyl)-benzene indicated that conduction did not occur through trimethylstannyl terminated molecules. In our past work[7], this was confirmed by showing that the conductance of auryltriphenylphosphine terminated compounds were the same as those terminated by SnMe$_3$ groups.

In addition to plateaus seen at a high conductance, the traces in Figure 2a show a second series of plateaus at 0.001 G$_0$ for **P1** (red) and $3\times10^{-5}$ G$_0$ for **P2** (blue). These are attributed to the *in situ* dimerization of the target compounds after the SnMe$_3$ linkers have been lost on the gold electrodes, in which the conjugated systems are linked by a dimethylene bridge. Indeed, some traces show two plateaus, one due to the monomer and one from the dimer. However, we cannot determine, based on conductance alone, whether both are present during the entire measurement. We will return to these features later in this paper.

Repeated measurements give a statistical assessment of the junction properties. In Figure 2b, we show conductance histograms for each of the compounds studied here. Each conductance histogram, generated from over 10000 traces without any data selection, reveals clear peaks at conductance values that depend on the molecular backbone. The inset of Figure 2b shows the same conductance histograms on a linear scale around 1 G$_0$. Here, we see a clear peak around 0.9 G$_0$ for **P1** that can be distinguished from the peak near 1 G$_0$ that is due to the conductance through a single gold-atom contact. By fitting the peaks in these conductance histograms with Lorentzians, we determine that the conductance of **P1** is about 0.9 G$_0$ while **P2**, **P3** and **P4** have a conductance of 0.1 G$_0$, 0.014 G$_0$ and 0.0022 G$_0$ respectively. The position of the highest conductance peak in each histogram is plotted on a semi-log scale against the number of phenyl rings in the molecule in Figure 2c. We find that for the series P2 to P4, the conductance decays exponentially with increasing number of phenyl groups with a decay constant $\beta$ = 1.9/phenyl or 0.43/Å. For comparison, in Figure 2c, we also plot the conductance of polyphenyls doubly-terminated with amine linkers, which show a similar decay in conductance with length.[12]

The histogram for **P1**, in Figure 2b, also shows a peak around 0.001 G$_0$ due to the presence of dimer molecules formed *in situ* during the measurement. To show that this is indeed due to conduction through a dimer molecule, we synthesized the di-tin precursor of p-xylylene dimer (**P1d**), and measured its conductance in the STM-BJ set-up (structure shown in Figure 3a). The conductance histogram for **P1d** shows a clear peak at 0.001 G$_0$ but no feature other than the gold conductance peak at 1 G$_0$ (See SI Figure 1). This clearly demonstrates that the dimer molecule **P1d** is created *in situ* when measurements of **P1SnMe$_3$** are carried out. That the conductance of the dimer is almost three orders of magnitude lower than the monomer can be attributed to the saturated dimethylene bridge between the two conjugated parts.

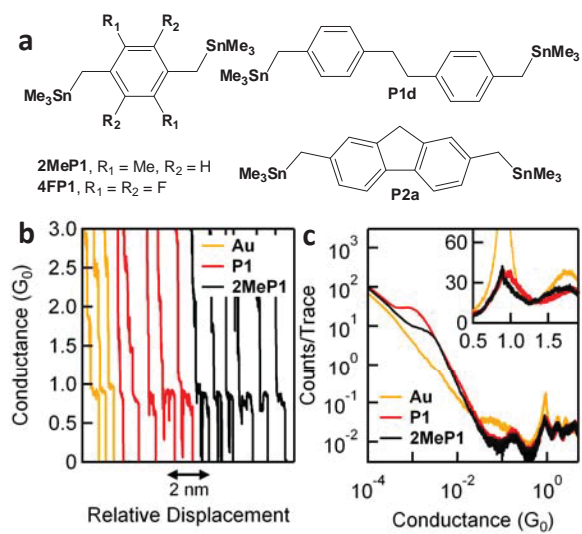

**Figure 3.** (**a**) Structures of additional compounds studied. (**b**) Individual conductance traces measured in solutions of the SnMe$_3$ terminated compounds **P1** (red) and **2MeP1** (black) at 25 mV applied bias. (**c**) Conductance histograms of over 10000 individual measurements generated with linear bin size of 0.0001 G$_0$ shown on a log-log scale. Inset: same histograms on a linear scale.

We also synthesized and measured the conductances of two xylylene derivatives with trimethyltin terminations (see SI for details). The first is a dimethyl-substituted xylylene (**2MeP1**) and the second is a tetrafluoro-substituted xylylene (**4FP1**) (structures shown in Figure 3a). For the dimethyl substituted xylylene (**2MeP1**), we see a conductance peak at around 0.9 G$_0$, very close to that of the unsubstituted **P1**, as shown in Figure 3c. The fact that the methyl substituents do not affect conductance significantly is consistent with a near-resonant transport mechanism, as will be discussed further below. Thus for for both **P1** and **2MeP1**, we demonstrate near-resonance transport across a molecular junction 0.8 nm in length.

The histogram for **2MeP1** in Figure 3c also shows a second peak around 0.002G$_0$, due to the formation of the dimer molecule, as in the case of **P1**. However, here we see a change in the conductance of the **P1** dimer when compared with the **2MeP1** dimer, as we expect when the mechanism for transport involves tunneling through the saturated ethano group. In contrast to the results with **2MeP1**, we find that the tetrafluoro-substituted xylylene (**4FP1**) does not show clear evidence for junction formation. The conductance histogram generated from 10000 measurements does not show two clear peaks around G$_0$ or a peak due to the molecular dimer formed *in situ* (see SI Figure 2). We attribute this to a stronger Sn-C bond in **4FP1SnMe$_3$**, making the *in situ* cleavage of the SnMe$_3$ group more difficult.

In our past work with Au-C coupled alkanes and benzene,[7] we found that the conductance of 1,4-didehydrobenzene covalently bonded to Au electrodes was only 0.03 G$_0$, significantly lower than that of p-xylylene. For benzene, only the molecular sigma system, which is a rather poor conductor, was well coupled through the Au-C bonds.[7-8] In contrast, with xylylene the Au-C bonds are very well coupled to the π system, yielding the high conductances observed. In principle, this coupling will depend on the angle between the Au-C bond and the phenyl plane and will be maximum when this angle is 90 degrees.[13] Based on calculations for this system we find indeed that the minimum energy configuration has a 90 degree angle, and the barrier for rotation is 10.4 kcal/mol or 0.45 eV (see SI Figure 4).

The conductance of **P2,** the biphenyl analog of **P1,** is 0.10 G$_0$ (Figure 2b). This conductance is a factor of 9 lower than that of **P1**, while in the case of amine-terminated polyphenyls, the differences between benzene and biphenyl was a factor of only ~6. To see if part of the origin of the difference between **P1** and **P2** is due to an anomalous internal twist angle at the central C-C bond, we synthesized a trimethyltin-terminated analog of fluorene with two methylene groups (**P2a**) and measured its conductance using the STM-BJ set-up. We find that **P2a** has a conductance of 0.17 G$_0$ (see SI Figure 3), but the ratio in conductance between **P1** and **P2a** (5.3) is still larger than the ratio for the diamine analogs (4.3).[9] This indicates that the lower conductance of an Au-**P2**-Au junction when compared with that of the Au-**P2a**-Au junction partly reflects a twist in the phenyl-phenyl bond that is absent in the fluorene analog. However, we see that an Au-**P1**-Au junction still has a higher conductance than one would expect, just extrapolating the exponential dependence seen for the longer polyphenyl compounds investigated here.

To understand the origin of the high conductance observed in these junctions, we carried out DFT-based first-principles calculations[14] with a gradient corrected exchange-correlation functional[15] and a non-equilibrium Green's function approach[16] to calculate electronic transmission through these junctions (see SI for details). Transmission curves for **P1, P2, P3** and **P4** junctions are shown in Figure 4. The calculated zero bias conductances are 0.9 G$_0$, 0.5 G$_0$, 0.15 G$_0$, and 0.05 G$_0$ for **P1**-**P4**, respectively. Transmission at the Fermi level is derived from the molecular orbitals (MOs) on the Au-C bonds that are very well coupled to the molecular π backbone and to the gold electrodes. For **P1**, this results in two distinct resonances, one for an even combination of the Au-C bond MOs and one for the odd combination as seen from the isosurface plots of the transmitted scattering state for each resonance (Figure 4a).[17] Physically, the Au-C bond MOs and the nearby π backbone MOs are nominally fully occupied by electrons. The extent to which the Fermi level for this junction falls within the nearest resonance depends on the amount of charge transfer from the molecule to the electrodes. An analysis of the Mulliken populations for the **P1** junction shows net positive charge, with the molecule having lost about 0.5 electrons. The electrostatic balance leads to the Fermi energy being placed slightly above the highest MO resonance resulting in conductance of **P1** being near resonant with a magnitude close to G$_0$.

For the longer derivatives, the effective through π-system coupling between the Au-C MOs is reduced. The corresponding energy splitting between the even and odd combinations of these MOs also gets smaller and the two distinct resonances seen for **P1** merge into a single, broad feature at around -0.5 eV, with decreased transmission at the peak (Figure 4b). However, the distinct even and odd combinations of the Au-C MOs can still be clearly seen in the transmitted scattering states (SI Figures 5-7). The charge transfer from the molecule to the electrodes is similar to **P1** and the Fermi level is pinned at an energy just above the highest resonance. The computed decay constant β for the **P2**, **P3** and **P4** is 1.2/phenyl group.

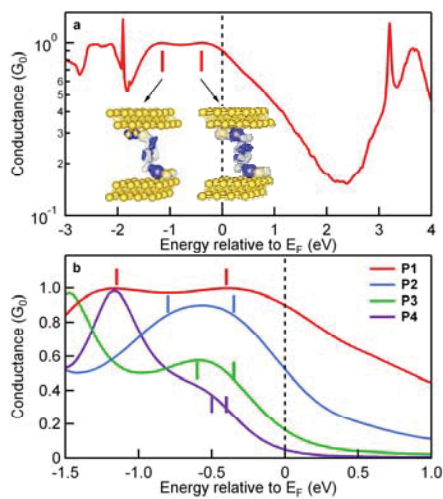

**Figure 4.** (a) Calculated transmission spectrum for **P1** bonded to Au electrodes. Inset: Isosurface plots of the real part of the transmitted scattering states for energies at the vertical bars (-1.15eV and -0.4 eV) showing the even and odd combinations of the Au-C bonds coupled through the π backbone. (b) Transmission spectra for **P1-P4**. Bars indicate the approximate position Au-C MO energies.

There are inherent errors in the use of the DFT molecular orbitals and energies for transport calculations in nanoscale junctions.[18] While the impact is minor for cases where the junction conductance is close to $G_0$, e.g. for metal point contacts[19], the calculated conductance values in the tunneling regime for single molecule junctions are typically larger than those measured in experiment.[18d, 18f, 20] In the present case, we expect that corrections to the DFT-based theory will only change the **P1** transmission modestly, leaving a resonance with near unit transmission close to the Fermi energy. However, the DFT-calculated π backbone MO energy is likely too close to the Fermi energy in general, an effect that will be larger for longer oligomers, where screening by the electrodes becomes less effective. In this case, the conductance will be smaller than indicated by the DFT calculations, with an increase in the effective beta value for **P2-P4**.

In conclusion, we have demonstrated a clear method to create circuits with strong electronic coupling between gold electrodes and conjugated molecules. We achieve a single molecule junction conductance close to one quantum across a length of approximately 0.8 nm. This remarkable result opens up new methods to create long and highly conducting molecular junctions.

## ASSOCIATED CONTENT

**Supporting Information available.** Synthetic procedures, measurement details, additional data and computational details. This material is available free of charge via the Internet at http://pubs.acs.org.

## AUTHOR INFORMATION

### Corresponding Authors

rb33@columbia.edu; mhyberts@bnl.gov; lv2117@columbia.edu

## ACKNOWLEDGMENT

This work was supported in part by the Nanoscale Science and Engineering Initiative of the NSF (award CHE-0641523), the New York State Office of Science, Technology, and Academic Research (NYSTAR) and NSF Career Award CHE-07-44185 (LV). JRW was supported by the EFRC program of the US Department of Energy (DOE) under Award Number DE-SC0001085 and carried out all STM-BJ measurements reported here. STS was the recipient of a Guthikonda Graduate Chemistry Fellowship. Part of this work was carried out at the Center for Functional Nanomaterials, Brookhaven National Laboratory, which is supported by the US Department of Energy, Office of Basic Energy Sciences, under contract number DE-AC02-98CH10886.